\documentclass[aps,prd,10pt,twocolumn, superscriptaddress, nofootinbib]{revtex4}
\usepackage[dvips]{graphicx}
\usepackage{epsfig, comment, caption}
\usepackage{multirow}
\usepackage{latexsym,amsmath,amsfonts,amssymb}
\usepackage{mathtools} 
\usepackage{slashed} 
\usepackage{blkarray}
\usepackage[all]{xy} 
\usepackage[makeroom]{cancel}
\usepackage[latin1]{inputenc}
\usepackage[american]{babel}
\usepackage{bbm}
\usepackage{color}
\usepackage[unicode]{hyperref}
\hypersetup{ linktoc=all,
    colorlinks, linkcolor={palatinateblue},
    citecolor={brightpink}, urlcolor={amaranth}
}

\graphicspath{{Images/}}

\definecolor{rosy}{RGB}{230,235,252}
\definecolor{myframetitle}{RGB}{90,89,170}
\definecolor{myblocktitle}{RGB}{140,185,249}
\definecolor{mytitle}{RGB}{10,80,26}

\definecolor{darkgreen}{RGB}{27,130,45}
\definecolor{darkblue}{rgb}{0,0,0.3}
\definecolor{darkred}{rgb}{0.7,0,0}

\definecolor{light gray}{RGB}{220,220,220}
\definecolor{dark purple}{RGB}{108,0,217}
\definecolor{pink}{RGB}{190,20,100}
\definecolor{orang}{RGB}{193,63,0}
\definecolor{green}{RGB}{11,98,17}
\definecolor{darkpink}{RGB}{153,0,76}
\definecolor{bluegreen}{RGB}{0,102,102}
\definecolor{greenlagan}{RGB}{0,102,0}
\definecolor{redgreen}{RGB}{102,102,0}
\definecolor{Redgreen}{RGB}{153,76,0}
\definecolor{vividviolet}{rgb}{0.62, 0.0, 1.0}
\definecolor{amaranth}{rgb}{0.9, 0.17, 0.31}
\definecolor{palatinateblue}{rgb}{0.15, 0.23, 0.89}
\definecolor{brightpink}{rgb}{1.0, 0.0, 0.5}
\definecolor{cornflowerblue}{rgb}{0.39, 0.58, 0.93}
\definecolor{deepcarminepink}{rgb}{0.94, 0.19, 0.22}
\definecolor{radicalred}{rgb}{1.0, 0.21, 0.37}

\def\6{\partial}

\newcommand{\be}{\begin{equation}}
\newcommand{\ee}{\end{equation}}
\newcommand{\beq}{\begin{equation}}
\newcommand{\eeq}{\end{equation}}
\newcommand{\bea}{\begin{eqnarray}}
\newcommand{\eea}{\end{eqnarray}}
\newcommand{\nn}{\nonumber \\}

\newcommand{\ba}{\begin{eqnarray}}
\newcommand{\ea}{\end{eqnarray}}

\newcommand{\beqs}{\begin{eqnarray}}
\newcommand{\eeqs}{\end{eqnarray}}
\newcommand{\bal}{\begin{aligned}}
\newcommand{\eal}{\end{aligned}}

\def\lbldef#1#2{\expandafter\gdef\csname #1\endcsname {#2}}

\def\href#1#2{#2}

\newcommand{\ber}{\begin{eqnarray}}
\newcommand{\eer}{\end{eqnarray}}

\newcommand{\beqar}{\begin{eqnarray}}

\newcommand{\eeqar}{\end{eqnarray}}

\newcommand{\dsl}
   {\kern.06em\hbox{\raise.15ex\hbox{$/$}\kern-.56em\hbox{$\partial$}}}

\newcommand{\eeqarr}{\end{eqnarray}}
\newcommand{\ZZ}{{\rm \kern 0.275em Z \kern -0.92em Z}\;}

\def\CC{{\mathchoice
{\rm C\mkern-8mu\vrule height1.45ex depth-.05ex
width.05em\mkern9mu\kern-.05em}
{\rm C\mkern-8mu\vrule height1.45ex depth-.05ex
width.05em\mkern9mu\kern-.05em}
{\rm C\mkern-8mu\vrule height1ex depth-.07ex
width.035em\mkern9mu\kern-.035em}
{\rm C\mkern-8mu\vrule height.65ex depth-.1ex
width.025em\mkern8mu\kern-.025em}}}

\def\RR{{\rm I\kern-1.6pt {\rm R}}}

\def\ZZ{{\rm Z}\kern-3.8pt {\rm Z} \kern2pt}
\def\IB{\relax{\rm I\kern-.18em B}}
\def\ID{\relax{\rm I\kern-.18em D}}
\def\II{\relax{\rm I\kern-.18em I}}
\def\IP{\relax{\rm I\kern-.18em P}}

\newcommand{\bear}{\begin{eqnarray}}
\newcommand{\eear}{\end{eqnarray}}

\def\6{\partial}

\newfont{\namefont}{cmr10}
\newfont{\addfont}{cmti7 scaled 1440}
\newfont{\boldmathfont}{cmbx10}
\newfont{\headfontb}{cmbx10 scaled 1728}

\newcommand{\dd}{\textrm{d}}

\allowdisplaybreaks[1]

\numberwithin{equation}{section}

\begin{document}

\title{Does Hubble Tension Signal a Breakdown in FLRW Cosmology?}


\author{C. Krishnan} \email{chethan.krishnan@gmail.com}
\affiliation{Center for High Energy Physics, Indian Institute of Science, Bangalore 560012, India}
\author{R. Mohayaee} \email{mohayaee@iap.fr}
\affiliation{Sorbonne Universit\'e, CNRS, Institut d'Astrophysique de Paris, 98bis Bld Arago, Paris 75014, France}
 \author{E. \'O Colg\'ain}\email{eoin@sogang.ac.kr}
 \affiliation{Center for Quantum Spacetime, Sogang University, Seoul 121-742, Korea}
 \affiliation{Department of Physics, Sogang University, Seoul 121-742, Korea} 
 \author{M. M. Sheikh-Jabbari}\email{shahin.s.jabbari@gmail.com}
\affiliation{School of Physics, Institute for Research in Fundamental Sciences (IPM), P.O.Box 19395-5531, Tehran, Iran}
\author{L. Yin}\email{yinlu@sogang.ac.kr}
\affiliation{Center for Quantum Spacetime, Sogang University, Seoul 121-742, Korea}
 \affiliation{Department of Physics, Sogang University, Seoul 121-742, Korea} 

\begin{abstract}
The tension between early and late Universe probes of the Hubble constant has motivated various new FLRW cosmologies. Here, we reanalyse the Hubble tension with a recent age of the Universe constraint. This allows us to restrict attention to matter and a dark energy sector that we treat without assuming a specific model. Assuming  analyticity of the Hubble parameter $H(z)$, and a generic low redshift modification to flat $\Lambda$CDM, we find that low redshift data ($z \lesssim 2.5$) and well-motivated priors only permit a dark energy sector close to the cosmological constant $\Lambda$. This restriction rules out late Universe modifications within FLRW. We show that early Universe physics that alters the sound horizon can yield an upper limit of $H_0 \sim 71 \pm 1$ km/s/Mpc. Since various local determinations may be converging to $H_0 \sim 73$ km/s/Mpc, a breakdown of the FLRW framework is a plausible resolution. We outline how future data, in particular strongly lensed quasar data, could also provide further confirmations of such a resolution. \\

\textit{``How often have I said to you that when you have eliminated the impossible, whatever remains, however improbable, must be the truth?" - Sherlock Holmes} 
\end{abstract}

\maketitle

\section{Introduction}
A cursory glance at Figure 1 of \cite{DiValentino:2021izs} reveals almost all  local $H_0$ determinations are biased higher than Planck-$\Lambda$CDM \cite{Aghanim:2018eyx}. This observation is agnostic about whether one calibrates Type Ia supernovae with Cepheids \cite{Freedman:2012ny, Riess:2016jrr, Riess:2019cxk, Riess:2020fzl} or TRGB \cite{Jang:2017dxn, Freedman:2019jwv, Freedman:2020dne, Soltis:2020gpl}. If true, this has profound consequences. In particular, one can argue that dark energy (DE), a feature in multiple datasets -  Type Ia supernovae \cite{Riess:1998cb, Perlmutter:1998np}, cosmic microwave background (CMB) \cite{Aghanim:2018eyx} and baryon acoustic oscillations (BAO) \cite{Eisenstein:2005su} - cannot be described by {a cosmological constant or by} a minimally coupled scalar field theory \cite{Vagnozzi:2018jhn, Colgain:2019joh, Banerjee:2020xcn}. Simply put, {\textit{uncoupled}} quintessence models \cite{Wetterich:1987fm, Ratra:1987rm} may already be  ruled out by the Hubble tension \cite{Banerjee:2020xcn}. {This may leave only \textit{coupled} quintessence models as viable options, e. g. \cite{Barros:2018efl, Gomez-Valent:2020mqn, SolaPeracaula:2020vpg}.} The key point is that Hubble tension appears to be extremely challenging for late time DE.  

Recall the age of Universe $t_{_U}$ within the Friedmann-Lema\^itre-Robertson-Walker (FLRW) paradigm: 
\be
\label{tU}
t_{_U} = \frac{977.8}{H_0} \int_{0}^{\infty} \frac{1}{(1+z) E(z)} \dd z \textrm{ Gyr}, 
\ee
where $H_0$ is the Hubble constant and $E(z)$ denotes the normalised Hubble parameter. Since $E(z)$ increases as $E(z) \sim (1+z)^p, p\geq 3/2$ in the early Universe,  the contribution of the early Universe physics to the integral is negligible. In other words, a competitive determination of $t_{_U}$ consistent with Planck, necessitates a late Universe modification to Planck-$\Lambda$CDM \cite{Bernal:2021yli}. 

Since the high redshift contribution to $t_{_U}$ is essentially fixed by $p=3/2$ and $\omega_{m} := \Omega_{m0} h^2$, which is in turn constrained by CMB, any alternative FLRW cosmology motivated by Hubble tension requires $H(z)$ higher than Planck at $z \approx 0 $, but lower at cosmological redshifts. This can be interpreted as a ``running of $H_0$" with redshift \cite{Krishnan:2020vaf}. More concretely, given two FLRW cosmologies, e.g. flat $\Lambda$CDM $H_{\tiny{\Lambda\textrm{CDM}}}(z)$ and a replacement $H_{\textrm{\tiny{new}}}(z)$, provided $H(z)$ is observationally determined, then $H_{0}^{\tiny{\Lambda\textrm{CDM}}}/H_{0}^{\textrm{\tiny{new}}} \sim E(z)_{\textrm{\tiny{new}}}/E(z)_{\tiny{\Lambda\textrm{CDM}}}$ must hold. Only when the two models are identical the left hand side can remain a constant. Such a running  must follow if Hubble tension can find a resolution within FLRW \cite{Krishnan:2020vaf}. 
Tentative hints of such a feature already exist \cite{Wong:2019kwg, Millon:2019slk, Yang:2019vgk, Krishnan:2020obg, Dainotti:2021pqg}.

In this letter, given a recent competitive determination for the Universe's age from globular clusters $t_{_U} = 13.5 \pm 0.27$ Gyr \cite{Bernal:2021yli, Valcin:2020vav}, we explore the possibility of the required late time modification, namely a higher $H_0$, with lower $E(z)$ at redshifts $z \lesssim 2.5$. To be as model independent as possible we employ overlapping Taylor expansions about small and large $z$. The first quantifies our ignorance of the DE sector, while the second is fixed by CMB determinations of $\omega_{m}$. We reduce the dependence of the latter on the DE sector by removing lower CMB multipoles $\ell \lesssim 30 $ \cite{Vonlanthen:2010cd, Audren:2012wb, Audren:2013nwa, Verde:2016wmz}. We also make use of a ``SH0ES prior" on the absolute magnitude of Type Ia supernovae $M_{B}$ \cite{Efstathiou:2021ocp} (see also \cite{Camarena:2021jlr}). {It is worth noting that a prior on $H_0$ is stronger than a prior on $M_B$, as the former affects all the cosmological data. Our use of a SH0ES prior, in contrast to a TRGB prior, is supported by a number of other independent local determinations that favour $H_0 > 70$ km/s/Mpc. These include megamasers \cite{Pesce:2020xfe}, surface brightness fluctuations \cite{Blakeslee:2021rqi} and Tully-Fisher relation \cite{Kourkchi:2020iyz}, which are either completely independent of the Cepheid distance ladder, or show little variation in $H_0$ across Cepheids and TRGB.}

This constitutes a minimal setup covering a matter sector and a late time deceleration-acceleration transition, without invoking any specific DE model. However, there is a loose end, namely the radius of the sound horizon $r_d$. When treated as a free parameter, we can allow putative early Universe physics that alters $r_d$ \cite{Poulin:2018cxd, Kreisch:2019yzn, Agrawal:2019lmo, Niedermann:2019olb, Niedermann:2020dwg}. We will also fix it through a prior \cite{Verde:2016wmz} and take the early Universe to be flat $\Lambda$CDM. {While our work overlaps with earlier studies \cite{Bernal:2016gxb, Lemos:2018smw},  our approach is analytic, it is relatively model-agnostic and leads to an upper bound on an achievable Hubble constant, $H_0 \sim 71 \pm 1$ km/s/Mpc, which represents a competitive 1.4\% error.}

Overall, we find that any freedom given to the DE sector in the late Universe is removed by the data and priors: our best-fit DE sector is more or less $\Lambda$. From this observation, results follow. First, the data \textit{must} adjust $r_d$ in any attempt to accommodate a SH0ES prior. Unsurprisingly, we find no running in $H_0$ and this disfavors  a resolution to Hubble tension in the DE sector $ z \lesssim 2.5$, but leaves an early Universe modification on the table for  $H_0 \sim 71$ km/s/Mpc or lower. Higher values, e. g. $H_0 \sim 73$ km/s/Mpc \cite{Riess:2020fzl}, are consistent within $2 \sigma$, but the discrepancy in central values allows room for other explanations. An alternative scenario is the breakdown of FLRW itself. In section \ref{sec:rabbithole} we argue  that $H_0$ values inferred from lensed quasars \cite{Wong:2019kwg, Millon:2019slk} may provide the first glimpse of the validity of this scenario.

\section{Setup} 
 It is assumed that the Universe is dominated by pressure-less dust that is well-approximated by the equation of state (EOS), $w = 0$. Furthermore, we have evidence \cite{Riess:1998cb, Perlmutter:1998np} that the Universe goes through a deceleration-acceleration transition at late times. Therefore, at low redshifts, the minimal model contains two sectors, namely DE and matter, with a Hubble parameter: 
\be
\label{Hubble}
H(z) = H_0 \sqrt{(1-\Omega_{m0} ) f(z) + \Omega_{m0} (1+z)^3}. 
\ee
Observe that other putative sectors currently constitute unmotivated speculation. 
As remarked in \cite{Krishnan:2020vaf}, $H_0$ is an integration constant and this fixes the overall scale in any FLRW cosmology. Moreover, the Friedmann equation also determines the RHS of (\ref{Hubble}) subject to the matter EOS, $w = 0$. Finally, the unknown function $f(z)$ describes the DE sector, but consistency demands that $f(z=0) = 1$. The flat $\Lambda$CDM model corresponds to $f(z) = 1$. 

\begin{figure}[htb]
\centering
  \includegraphics[width=75mm]{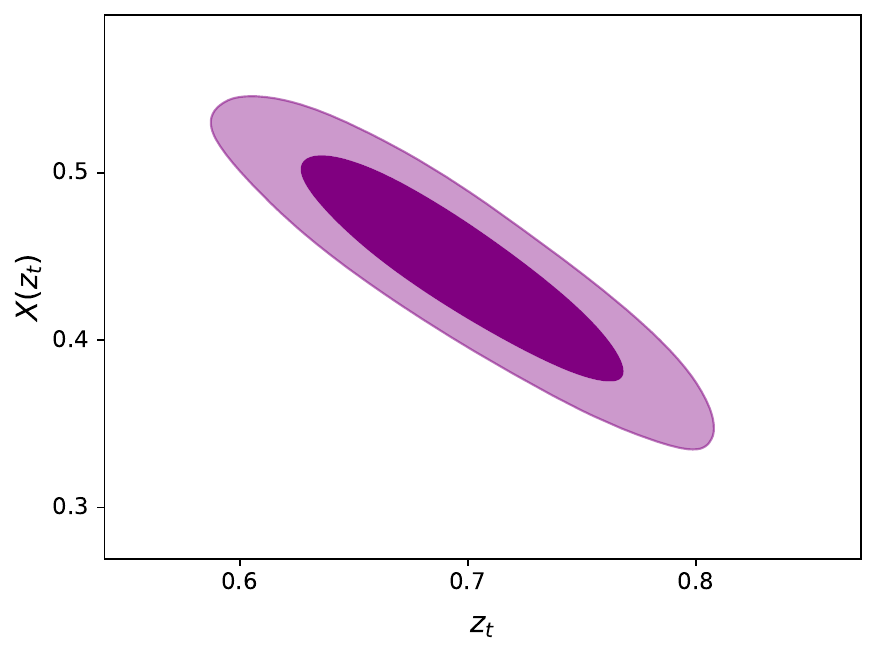}
\caption{Distribution of the transition redshift $z_t$ and $X(z_t)$ inferred from Planck MCMC chains of CMB+BAO+supernovae constraints on the CPL model.}
\label{Xzt}
\end{figure}

{To study  DE contribution it is convenient to write $H(z)$ as},\footnote{There are strong constraints on ``early dark energy'' \cite{Wetterich:2004pv, Doran:2006kp} from CMB and BBN \cite{Bean:2001wt, Pettorino:2013ia, Ade:2015rim}, but one can still play with the idea pre-recombination \cite{Poulin:2018cxd, Niedermann:2019olb, Niedermann:2020dwg}.} 
\bea
\label{H}
 H(z) &=& H_0 \sqrt{\Omega_{m0}} (1+z)^{\frac{3}{2}} \sqrt{ 1+ X},
 \eea
 \bea
\label{X}
X &:=& \frac{\Omega_{\Lambda}(z)}{\Omega_{m}(z)} =\frac{ (1- \Omega_{m0} ) f(z)}{\Omega_{m0} (1+z)^3}.
\eea
Now, if $X(z)$ is small and decreasing $X'(z) < 0$, we can treat it perturbatively at higher redshifts. Recall that by definition $ \ddot{a}/a = \dot{H} + H^2$. As a result, when one moves from deceleration ($\ddot{a} < 0$) to acceleration ($\ddot{a} > 0$), necessarily $\dot{H} + H^2 = 0$ at the transition redshift $z_t$. We can get an idea for the size of $X(z_t)$ in the Chevallier-Polarski-Linder (CPL) parametrisation \cite{Chevallier:2000qy, Linder:2002et}, where concretely 
\be
\label{CPL}
f(z)  = (1+z)^{3 (1+w_0 + w_a)} e^{-\frac{3 w_a z}{1+z}}. 
\ee
To do this, one imports Planck Markov Chain Monte Carlo (MCMC) chains \cite{Aghanim:2018eyx} and solves for $\ddot{a} = 0$ to identify $z_t$ for each entry in the MCMC chain. One can see from Figure \ref{Xzt}, where we have used \textit{getdist} \cite{Lewis:2019xzd}, that in relaxing the DE model away from $\Lambda$, the $(z_t, X(z_t))$  values are still close to the Planck-$\Lambda$CDM values, $(z_t, X(z_t)) \approx (0.67, 0.5)$. The key point is that $X(z>z_t)< X(z_t) < 1$, so we can treat it perturbatively. \footnote{That $z_t$ and $X(z_t)$ do not change much within the CPL model may not be so surprising, since CPL is biased towards $\Lambda$ \cite{Colgain:2021pmf}.}  

\section{Strategy}
Following \cite{Banerjee:2020xcn} we work perturbatively at low redshift by Taylor expanding $H(z)$ about $z = 0$. It is well documented that one cannot go beyond $z = 1$ \cite{Cattoen:2007sk} (see also Appendix A of \cite{Colgain:2021ngq}), so here we expand up to an intermediate redshift $z_{*} \sim z_t$.  Now, we can Taylor expand the unknown function $f(z)$, 
\be
\label{f}
f(z) = 1 + f_1 z + \frac{f_2}{2!}  z^2 + \frac{f_3}{3!} z^3 + \dots  
\ee
where we have defined $f_n \equiv f^{(n)}(0)$. Adopting this expansion, $H(z)$ for $z<z_*$ becomes, 
\be
\label{lowz}
H(z) = H_0 \left( 1 + h_1 z + h_2 z^2 + h_3 z^3 + \dots  \right),  
\ee
where we have further defined 
\begin{widetext}
\bea
h_1 &=& \frac{3}{2} \Omega_{m0} +  \frac{1}{2} (1- \Omega_{m0} ) f_1, \quad
h_2 =  \frac{3}{2} \Omega_{m0} - \frac{1}{8} ( 3 \Omega_{m0} + (1- \Omega_{m0}) f_1)^2 + \frac{1}{4} (1- \Omega_{m0} ) f_2, \nn
h_3 &=& - \frac{1}{4} ( 3 \Omega_{m0}+ (1- \Omega_{m0}) f_1) \left(3 \Omega_{m0} - \frac{1}{4} ( 3 \Omega_{m0} + (1 - \Omega_{m0} )f_1)^2 + \frac{1}{2} ( 1 - \Omega_{m0}) f_2 \right) \nn 
&+& \frac{1}{2} ( \Omega_{m0}  + \frac{1}{6} ( 1 - \Omega_{m0}) f_3). 
\eea
\end{widetext}
We restrict the presentation to third order as it  approximates flat $\Lambda$CDM well to $z_t$. Nevertheless, 
$f_3$ is degenerate with $\Omega_{m0}$, so one can safely set $f_3 = 0$.\footnote{Our expansion covers interacting DE models which can be written as $w(z)$CDM, e. g. \cite{Gavela:2009cy}. However, {curvature $\Omega_{k}$ is strictly speaking not covered by our definitions of distances, e. g. $d_{L}(z), d_{A}(z)$. More precisely, curvature can be absorbed into $f_i$ in $H(z)$, but it affects distances. However, this change will not be significant if the curvature is small, $\Omega_{k} \approx 0$.} } 

{To appreciate this fact, observe that the term in the square root in (\ref{H}) can be expanded to third order in $z$ as 
\bea
&& (1- \Omega_{m0}) f(z) + \Omega_{m0} (1+z)^3 \nn
&=& 1 + (1- \Omega_{m0}) \left(f_1 z + \frac{f_2}{2!} z^2 + \frac{f_3}{3!} z^3 \right) \nn
&+& \Omega_{m0} (3 z + 3 z^3 + z^3). 
\eea
However, one can absorb $f_3$ into the definition of a new $\Omega^{\prime}_{m0}$, while also redefining $f_1$ and $f_2$: 
\bea
\Omega^{\prime}_{m0} &=& \Omega_{m0} + \frac{f_3}{3!} (1- \Omega_{m0} ), \\
f_2^{\prime} &=& \frac{6 (f_3 - f_2)}{f_3 - 6}, \quad f_1^{\prime} = \frac{3(f_3-2 f_1)}{f_3 - 6}. \nonumber
\eea
The end result is the original expression with $f_3 = 0$. This degeneracy between $\Omega_{m0}$ and $f_3$ allows one to set $f_3 = 0$ in (\ref{lowz}). }  

We assume that (\ref{f}) is a valid perturbative expression in the sense that higher order terms are smaller, i. e. $  |f_n| z <  n |f_{n-1}|$ in the redshift range of interest $z < z_*$. Recalling that a number of late-time resolutions to Hubble tension correspond at the background level to a $w$CDM model with $w \approx -1.2$ \cite{Vagnozzi:2019ezj} (see also \cite{Li:2019yem, DeFelice:2020sdq, Heisenberg:2020xak}), it is useful to note the expansion of (\ref{CPL}) with representative values: $f(z)_{w = -1.2} = 1 - 0.6 z + 0.48 z^2+ O(z^3)$. 

The idea now is to use (\ref{lowz}) below $ z= z_*$, but to use the following expression for $z \geq z_*$: 
\be
\label{highz}
H(z) = H_0 \sqrt{\Omega_{m 0}} (1+z)^{\frac{3}{2}}  \left(1 + \frac{1}{2} X - \frac{1}{8} X^2 + \dots \right), \nn
\ee
where throughout we expand both $H(z)$ and $H(z)^{-1}$ to $O(X^{10})$. In line with expectations, the low redshift expansion (\ref{lowz}) starts to deviate from the exact result before $z = 1$, whereas the high redshift expansion can be pushed close, but not as far as $z = 0$. The low redshift expansion performs worse, yet the error is within $1 \%$ for $\Omega_{m0} = 0.3$ through to $z_{*} \sim 0.8$. In contrast, the error in the high redshift expansion is negligible. Note that $z_*$ constitutes a redundancy in our analysis, but as we show later (Table \ref{table2}), it does not change our results. 

\section{Data Sets}

We make use of the Pantheon supernovae dataset \cite{Scolnic:2017caz}, a rich compilation of BAO data \cite{Beutler:2011hx, Ross:2014qpa, Alam:2016hwk, Abbott:2017wcz, Bautista:2020ahg, Gil-Marin:2020bct, Raichoor:2020vio, deMattia:2020fkb, Hou:2020rse, Neveux:2020voa, duMasdesBourboux:2020pck}, as presented in Table \ref{table3}, and cosmic chronometer data \cite{Moresco:2016mzx}. 
To flesh out the analysis we employ of a number of priors. First, we assume that $t_{_U} = 13.50 \pm 0.27$ Gyrs \cite{Bernal:2021yli}. Next, we employ the high redshift prior $\omega_{m} = 0.141 \pm 0.006 $, where the lower multipoles have been removed to reduce dependence on the DE sector (see appendix \ref{sec:prior}). Finally, we utilise a SH0ES prior on the absolute magnitude, $M_{B} = -19.244 \pm 0.042$ mag \cite{Efstathiou:2021ocp}, which is consistent with \cite{Camarena:2021jlr}, and an early Universe prior $r_d = 147.4 \pm 0.7$ \cite{Verde:2016wmz}.  {Note that we adapt a prior on $M_B$ and a $H_0$ prior would be stronger than an $M_{B}$ prior.}

\begin{table}[htb]
\centering 
\begin{tabular}{c|c|c|c}
 & $z_{\textrm{eff}}$ & Constraint & Ref. \\
\hline
\rule{0pt}{3ex} 6dF & $0.106$ & $D_{V}/r_d = 2.98 \pm 0.13$ & \cite{Beutler:2011hx} \\
\hline
\rule{0pt}{3ex} SDSS MGS & $0.15$ & $D_{V}/r_d = 4.47 \pm 0.17$ & \cite{Ross:2014qpa} \\
\hline
\rule{0pt}{3ex} \multirow{4}{*}{BOSS Galaxy} &  \multirow{2}{*}{$0.38$} & $D_{A}/r_d = 7.41 \pm 0.12$ & \multirow{2}{*} {\cite{Alam:2016hwk}} \\
\rule{0pt}{3ex}  & & $D_{H}/r_d = 25.00 \pm 0.76$ &  \\
\cline{2-4}
\rule{0pt}{3ex}  &  \multirow{2}{*}{$0.51$} & $D_{A}/r_d = 8.85 \pm 0.14$ &  \multirow{2}{*} {\cite{Alam:2016hwk}} \\
\rule{0pt}{3ex}  & & $D_{H}/r_d = 22.38 \pm 0.58$ &  \\
\hline
\rule{0pt}{3ex} DES & $0.81$ & $D_{A}/r_d = 10.75 \pm 0.43$ & \cite{Abbott:2017wcz} \\
\hline
\rule{0pt}{3ex} \multirow{2}{*}{eBOSS LRG} &  \multirow{2}{*}{$0.70$} & $D_{A}/r_d = 10.51 \pm 0.19$ & \multirow{2}{*} {\cite{Bautista:2020ahg, Gil-Marin:2020bct}} \\
\rule{0pt}{3ex}  & & $D_{H}/r_d = 19.33 \pm 0.53$ &  \\
\hline
\rule{0pt}{3ex} eBOSS ELG &  $0.85$ & $D_{V}/r_d = 18.33^{+0.57}_{-0.62}$ &  \cite{Raichoor:2020vio, deMattia:2020fkb} \\
\hline
\rule{0pt}{3ex}  \multirow{2}{*}{eBOSS Quasar} &  \multirow{2}{*}{$1.48$} & $D_{A}/r_d = 12.38 \pm 0.32$ &  \multirow{2}{*} {\cite{Hou:2020rse, Neveux:2020voa}} \\
\rule{0pt}{3ex}  & & $D_{H}/r_d = 13.26 \pm 0.55$ &  \\
\hline
\rule{0pt}{3ex}  \multirow{2}{*}{Ly$\alpha$-Ly$\alpha$} &  \multirow{2}{*}{$2.33$} & $D_{A}/r_d = 11.29 \pm 0.57$ &  \multirow{2}{*}{\cite{duMasdesBourboux:2020pck}} \\
\rule{0pt}{3ex}  & & $D_{H}/r_d = 8.93 \pm 0.28$ &  \\
\hline
\rule{0pt}{3ex}   \multirow{2}{*}{Ly$\alpha$-Quasar} &  \multirow{2}{*}{$2.33$} & $D_{A}/r_d = 11.20 \pm 0.51$ &  \multirow{2}{*}{\cite{duMasdesBourboux:2020pck}} \\
\rule{0pt}{3ex}  & & $D_{H}/r_d = 9.08  \pm 0.34$ &  \\
\end{tabular} 
\caption{BAO data}
\label{table3}
\end{table}

\section{Results}
Consider first the flat $\Lambda$CDM model subject to the $t_{_U}$ prior and $r_d, M_B$ treated as nuisance parameters. The best-fit values are shown in Table \ref{table1}. Evidently, $H_0$ is higher than Planck \cite{Aghanim:2018eyx}, but this is not so surprising since we have adopted a $t_{_U}$ prior. From our MCMC chains, where we used \textit{emcee} \cite{ForemanMackey:2012ig}, we find that $t_{_U} = 13.62 \pm 0.16$ Gyr, while $\omega_{m} = 0.143 \pm 0.004$, which are both consistent with the priors. The higher $H_0$ appears to be driven by the $t_{_U}$ prior, while the nuisance parameters $r_d$ and $M_B$ come along for the ride. Nevertheless, $r_d$ is consistent with the Planck value $r_{d} \sim 147$ Mpc at $1 \sigma$. 

\begin{table}[htb]
\centering
\begin{tabular}{c|c|c|c}
\rule{0pt}{3ex} $H_0$ (km/s/Mpc) & $\Omega_{m0}$ & $r_d$ (Mpc) & $M_B$ (mag) \\
\hline
\rule{0pt}{3ex} $69.56^{+1.25}_{-1.23}$ & $0.295^{+0.013}_{-0.012}$ & $144.8^{+2.2}_{-2.0}$ & $-19.37^{+0.03}_{-0.03}$ 
\end{tabular}
\caption{Best fit values for the flat $\Lambda$CDM model with an $t_{_U}$ prior and a low multipole-subtracted prior on $\omega_{m}$. }
\label{table1}
\end{table}

The flat $\Lambda$CDM fit simply serves to get oriented. Next, we repeat, but now let the DE sector vary through the $f_i$ (\ref{f}) and impose a SH0ES prior on $M_{B}$ \cite{Efstathiou:2021ocp}. The best-fit values can be found in Table \ref{table2}, where we record results both with and without a prior on $r_d$ \cite{Verde:2016wmz}. Observe also that we have scanned over $z_{*}$, and as expected, it does not change our results. Throughout we find that $H_0$, $\Omega_{m0}$ and $t_{_U}$ are all fully consistent with the imposed priors.

\begin{widetext}
\begin{center} 
\begin{table}[htb]
\begin{tabular}{c|c|c|c|c|c|c}
\rule{0pt}{3ex} $z_{*}$ & $H_0$ (km/s/Mpc) & $\Omega_{m0}$ & $f_1$ & $f_2$ & $r_d$ (Mpc) & $M_B$ (mag) \\
\hline
\rule{0pt}{3ex} \multirow{2}{*}{$0.5$} & $70.97^{+1.02}_{-1.00}$ & $0.284^{+0.012}_{-0.010}$ & $0.15^{+0.13}_{-0.17}$ & $-0.08^{+0.25}_{-0.32}$ & $141.88^{+1.8}_{-1.9}$ & $-19.32^{+0.03}_{-0.03}$ \\
\rule{0pt}{3ex} & $69.34^{+0.83}_{-0.79}$ & $0.292^{+0.013}_{-0.012}$ & $0.08^{+0.14}_{-0.22}$ & $-0.10^{+0.26}_{-0.32}$ & $146.8^{+0.7}_{-0.6}$ & $-19.37^{+0.02}_{-0.02}$ \\
\hline
\rule{0pt}{3ex} \multirow{2}{*}{$0.6$} & $71.18^{+1.04}_{-1.04}$ & $0.279^{+0.013}_{-0.013}$ & $0.17^{+0.13}_{-0.13}$ & $-0.02^{+0.25}_{-0.29}$ & $141.3^{+2.0}_{-1.9}$ & $-19.31^{+0.03}_{-0.03}$ \\
\rule{0pt}{3ex} & $69.39^{+0.85}_{-0.83}$ & $0.291^{+0.013}_{-0.013}$ & $0.07^{+0.15}_{-0.23}$ & $-0.06^{+0.25}_{-0.30}$ & $146.7^{+0.7}_{-0.7}$ & $-19.37^{+0.02}_{-0.02}$  \\
\hline 
\rule{0pt}{3ex} \multirow{2}{*}{$0.7$} & $71.15^{+1.03}_{-1.03}$ & $0.279^{+0.013}_{-0.013}$ & $0.18^{+0.13}_{-0.12}$ & $-0.03^{+0.22}_{-0.28}$ & $141.2^{+2.0}_{-1.8}$ & $-19.31^{+0.03}_{-0.03}$ \\
\rule{0pt}{3ex} & $69.38^{+0.84}_{-0.84}$ & $0.290^{+0.013}_{-0.012}$ & $0.07^{+0.15}_{-0.23}$ & $-0.05^{+0.22}_{-0.27}$ & $146.7^{+0.6}_{-0.6}$ & $-19.37^{+0.02}_{-0.02}$  \\
\end{tabular}
\caption{Best fit values for our overlapping expansions with a $t_{_U}$ prior, low multipole-subtracted prior on $\omega_{m}$ and a SH0ES prior on $M_B$. In the lower entries for each $z_*$ we impose a prior on $r_d$.}
\label{table2}
\end{table}
\end{center} 
\end{widetext} 

It is worth noting that with the $r_d$ prior, both $f_1$ and $f_2$ are consistent with their flat $\Lambda$CDM values within $1 \sigma$. There is clearly no freedom in the DE sector. The likely explanation for this is that our prior on $\omega_{m}$ leaves little freedom for non-zero $f_i$ within the constraints imposed by BAO and supernovae on $H(z)$. Moreover, observe that the constraints on $r_d$ and $t_{_U}$ lead to a value of $M_B$ that is the same as the flat $\Lambda$CDM fit (Table \ref{table1}). Tellingly, the final $M_B$ value is $2.7 \sigma$ away from the original prior and this underscores the tension between $M_{B}$ and $r_d$.

\begin{figure}[h]
\centering
  \includegraphics[width=75mm]{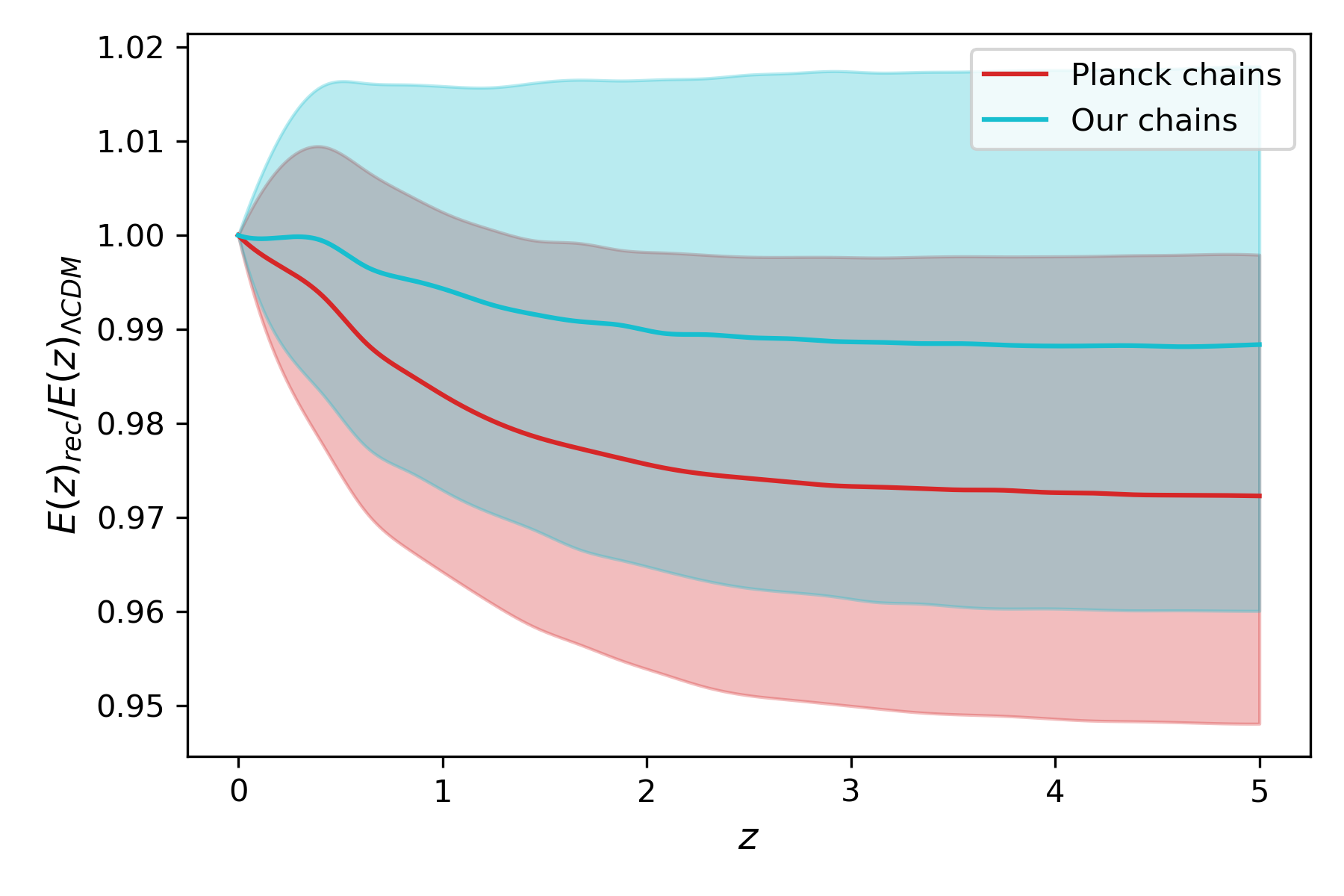}
\caption{The ratio between the normalised Hubble parameters for our cosmological model (\ref{Hubble}) and flat $\Lambda$CDM with a prior on $r_d$ and $z_* = 0.6$. {The shaded blue and red envelopes are the $1 \sigma$ confidence intervals, while the grey region is the overlap.} That this ratio is basically a constant confirms there is no $H_0$ running in the sense discussed in \cite{Krishnan:2020vaf}.}
\label{prior}
\end{figure}

In Figure \ref{prior} we offer visual confirmation that the reconstructed $E(z)_{\textrm{rec}}$ does not differ enough from flat $\Lambda$CDM $E(z)_{\Lambda\textrm{CDM}}$. There is no evidence for running in $H_0$ with redshift \cite{Krishnan:2020vaf} as is clear from the $1 \sigma$ confidence intervals for the ratio presented. For completeness, we compare to flat $\Lambda$CDM defined with respect to our MCMC chains (Table \ref{table1}) and the Planck MCMC chains \cite{Aghanim:2018eyx}. The asymptotic, large $z$ behaviour is governed by the best-fit values of $\Omega_{m0}$.

\begin{figure}[h]
\centering
  \includegraphics[width=75mm]{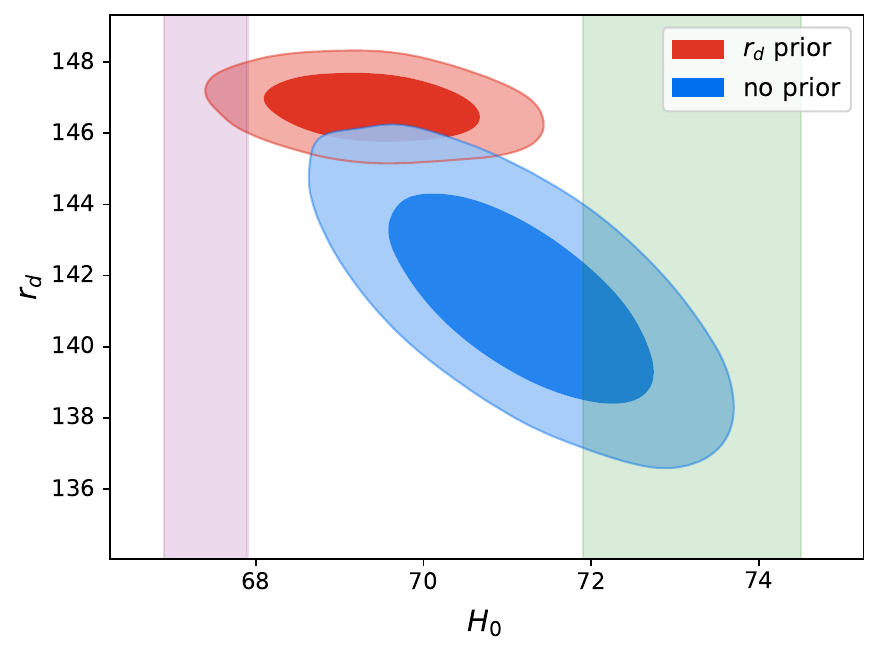}
\caption{Best-fit values of $H_0$ with and without a prior on $r_d$. The purple and green bands represents $1 \sigma$ from Planck and SH0ES \cite{Riess:2020fzl}, respectively. }
\label{H0rd}
\end{figure}

Finally, when $r_d$ is treated as a free parameter, $f_2$ is consistent with zero, but $f_1$ shows some marginal departure, which may be described by the simplest extension, the $w$CDM model. In essence, there may be some slight deviation from $\Lambda$ in the DE sector, as suggested in \cite{Bernal:2021yli}. However, {the maximum $H_0$ value achievable within an FLRW cosmology coupled to Einstein gravity, regardless of the DE sector, is $H_0 \sim 71 \pm 1 $ km/s/Mpc}. {We illustrate the difference the prior makes in Figure \ref{H0rd}, where the tension between SH0ES and $r_d$ with a flat $\Lambda$CDM prior is evident. The figure is plotted for $z_* = 0.6$, but $z_*$ does not affect the results (see Table \ref{table2}). This figure, although at low significance, clearly demonstrates that the early Universe solutions within FLRW come short of solving the Hubble tension.} 

\section{Going beyond FLRW}
\label{sec:rabbithole}
H0LiCOW \cite{Wong:2019kwg} has reported a descending trend of $H_0$ with lens redshift in the flat $\Lambda$CDM model and an extra time delay measurement was also obtained from DES \cite{2020MNRAS.494.6072S}.
 This is interesting since all lenses should be subject to the same assumptions, so if there is a systematic, then all lenses are affected. Nevertheless, the descending trend in $H_0$ can be expected to be robust \cite{Millon:2019slk} and it has prompted various explanations \cite{Krishnan:2020vaf, Haslbauer:2020xaa}. However, it is possible that it is not a trend with lens redshift, but due to the line of sight (LOS) anisotropy as we outline here.

Although the effect of local anisotropies on lensing is marginal, if these anisotropies extend to a far larger extent than predicted by LCDM model, as shown by bulk flow data, then they could have non-negligible impact on the lensing analyses. 
Indeed, a large over-density along the LOS will increase the estimated value of $H_0$   (by increasing the value of $\kappa_{ext}$, see
{\it e.g.} \cite{2010ApJ...711..201S} for details).

\begin{table}[h]
\centering
\begin{tabular}{c|c|c|c}
\rule{0pt}{3ex} Lens & $\alpha$ & $\delta$ & $H_0$ (km/s/Mpc) \\
\hline 
\rule{0pt}{3ex} B1608+656 & $242^{\circ}$ & $+66^{\circ}$ & $71.0^{+2.9}_{-3.3}$ \\
\rule{0pt}{3ex} RXJ1131-1231 & $173^{\circ}$ &$-13^{\circ}$ & $78.2^{+3.4}_{-3.4}$ \\
\rule{0pt}{3ex} HE 0435-1223 &  $70^{\circ}$ & $-12^{\circ}$& $71.7^{+4.8}_{-4.5}$ \\
\rule{0pt}{3ex} SDSS 1206+4332 & $182^{\circ}$ &$+44^{\circ}$ & $68.9^{+5.4}_{-5.1}$ \\
\rule{0pt}{3ex} WFI2033-4723 & $308^{\circ}$ & $-47^{\circ}$ & $71.6^{+3.8}_{-4.9}$ \\
\rule{0pt}{3ex} PG 1115+080 & $170^{\circ}$ & $+8^{\circ}$ & $81.1^{+8.0}_{-7.1}$ \\
\rule{0pt}{3ex} DES J0408-5354 & $62^{\circ}$ & $-54^{\circ}$ & $74.2^{+2.7}_{-3.0}$ \\
\end{tabular}
\caption{Location of lenses on the celestial sphere from H0LiCOW and DES (last data point).}
\label{table4}
\end{table}

\begin{figure}[htb]
  \includegraphics[width=76mm]{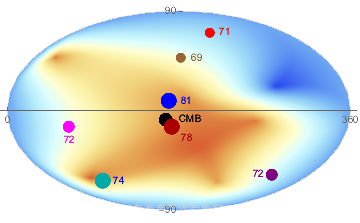}
\caption{\small{Sky position, in RA and DEC, of the lenses from the H0LiCOW sample and DES are shown on an Aitoff plot together with the direction of the CMB dipole. The inferred values of $H_0$ for each lens are shown next to the lens position. The data points are overlaid on a density plot of mass distribution in the local Universe (Virgo, Great Attractor, Shapley, Ophiuchus, Coma, Pavo-Indus, Hercule, Norma,
 Perseus-Pisces, Horologium and the local void are included). The large DES value is near the Horologium supercluster and the two large ones from H0LiCOW are near most other superclusters including Shapley. The supercluster on the top left is  Perseus-Pisces. The dark blue region on the top right quadrant is the local void. GP is not shown.}}
\label{H0LiCOW-sky}
\end{figure}

To begin, instead of {focusing on the dependence of $H_0$ values on lens redshift, as is done in \cite{Wong:2019kwg, Millon:2019slk}, we study their angular distribution on the sky.} 
In Table \ref{table4} we present the coordinates of the H0LiCOW and DES lenses in the equatorial coordinate system where $\alpha$ and $\delta$ denote right ascension (RA) and declination (DEC), respectively. It is intriguing to compare the orientations of these few quasars with the CMB dipole \cite{Aghanim:2018eyx},  $\alpha = 168^{\circ}, \quad \delta = -7^{\circ}, $ approximately along which most of the superclusters in the local Universe are lined up. This anisotropy, if extended to larger scales as shown by non-convergence of the CMB dipole, is expected to effect the LOS contribution to the time-delay.
The LOS contribution has been taken into account within certain models, but to our knowledge the effect of anisotropies or the aberration effect due to large velocity of the observer with respect to distant quasars has not been reported in these studies.
Although we currently only have a tiny dataset of seven lensed quasars, we clearly see that the two lenses that are closest to the direction of the CMB dipole, namely RXJ1131-1231 and PG 1115+080, exhibit the highest values of $H_0$. Furthermore, a large value of $H_0$ is inferred from DES data, which lies close to the direction of the Horologium supercluster. This may not be a coincidence or a simple fluctuation. A similar study has been carried out on the anisotropy of $H_0$ using supernovae, which due to poor data yielded a low-significant result \cite{2010MNRAS.401.1409C}.
Clearly these data are significantly impacted by the degeneracy in the mass distribution modelling of the lenses, however future similar data on time delay from LSST shall certainly provide us with further confirmations.

It is often assumed that the origin of the CMB dipole is the gravitational attraction of the local superclusters. However it seems that the observed superclusters up to the Shapley concentration cannot fully account for the CMB dipole, which consequently cannot be of purely kinematic origin ({\it e.g.} see \cite{Colin:2018ghy} and references therein).
Furthermore, since the sources in this dataset are all at high redshifts, then they might signal a large structural clustering well beyond what is permitted in an FLRW Universe. It is thus expected that the seemingly far-reaching local anisotropy has some signature on the strongly lensed quasars and this indeed appears to be the case. The correlation between $H_0$ from the lensed quasars and distribution of 
most-prominent superclusters is suggestive even for such a tiny dataset and deserves further study as data improves.

{We make one last comment before closing the section. We emphasise again that it is the relative differences in $H_0$ that interest us from the lensed quasars. In order to read into the correlations in Figure \ref{H0LiCOW-sky}, it is imperative that all lenses have been modeled uniformly. As is clear from recent results \cite{Birrer:2020tax}, the \textit{absolute} value of $H_0$ inferred from strong lensing time delay depends on assumptions for mass sheet modeling, so for this reason we have not employed it as a prior in our analysis.}

\section{Discussion}
In this letter we have explored a modification of the DE sector consistent with a recent determination of the age of Universe, $t_{_U}$  \cite{Bernal:2021yli, Valcin:2020vav}. As explained, a fixed $t_{_U}$ consistent with Planck results allows a higher $H_0$, but lower $E(z)$ relative to Planck-$\Lambda$CDM, thus providing a realisation of a running $H_0$ \cite{Krishnan:2020vaf}. Ultimately, the rigidity of the DE sector precludes this. As a result, data has a preference for an early Universe resolution, but only as high as $H_0 \sim 71$ km/s/Mpc. Of course, this assumes a working model can be found, ideally something natural that penalises additional unobserved degrees of freedom, (see {\it e.g.} \cite{Jedamzik:2020krr}), which is currently far from clear \cite{Hill:2020osr,Ivanov:2020ril,DAmico:2020ods, Niedermann:2020qbw, Murgia:2020ryi, Smith:2020rxx, Jedamzik:2020zmd, Lin:2021sfs, Vagnozzi:2021gjh}. Note, our analysis in principle captures all early Universe resolutions, since $r_d$ is a free parameter.  

What is new here is that our analysis is minimal, yet covers  $(w_0, w_a)$ parametrisations of DE (see \cite{Yang:2021flj}), while the $H_0$ errors are meaningful. As is clear from Table \ref{table2}, we have a 1.4$\%$ error on $H_0 \sim 71$ km/s/Mpc without assuming a prior on $r_d$. While this is only $1.2 \sigma$ discrepant with SH0ES \cite{Riess:2020fzl}, once local determinations are combined, for example \cite{DiValentino:2020vnx} or \cite{Verde:2019ivm}, this inches up to $1.4 \sigma$ and $1.7 \sigma$, respectively. Putting aside caveats about combining data, admittedly, even as data improves, it is hard to imagine any better than a $\sim 2 \sigma$ discrepancy with SH0ES. Of course, if claims of new early Universe physics are challenged by forthcoming results from ACT \cite{Aiola:2020azj}, SPT-3G \cite{Benson:2014qhw}, Simons Observatory \cite{Ade:2018sbj} and CMB-S4 \cite{Abazajian:2019eic}, we remain with the possibility of deviations from FLRW in the late Universe. Obviously, this deviation at late times should be considered with respect to a perturbed FLRW model. In contrast, when $r_d$ is fixed, the resulting $H_0$ is discrepant at $2.5 \sigma$ with SH0ES \cite{Riess:2020fzl} and in excess of $3 \sigma$ with local $H_0$ combinations \cite{DiValentino:2020vnx, Verde:2019ivm}. Late Universe modifications of the DE sector are firmly ruled out. 

Throughout it should be kept in mind that FLRW is a working assumption. Given the CMB is more or less isotropic, FLRW is well motivated in the early Universe.\footnote{Note that while there exist anomalies  ({see \cite{Schwarz:2015cma, Perivolaropoulos:2021jda} for reviews}) we cut out low $\ell$ multipoles, and hence the corresponding anisotropies, in our analysis.} The $4.9 \sigma$-significant result of \cite{Secrest:2020has}, which employs almost 1.5 million quasars primarily at $z>1$ implies that the CMB restframe does not coincide with the restframe of distant quasars, thus directly undermining the FLRW Universe. Similar results were noted earlier, at lower significance for radio galaxies at lower redshifts and for smaller samples \cite{Blake:2002gx, Singal:2011dy, Gibelyou:2012ri, Rubart:2013tx, Tiwari:2015tba, Colin:2017juj, Bengaly:2017slg, Siewert:2020krp}. Intriguingly, as we have explained in section \ref{sec:rabbithole}, strong lensing time delay may provide a new way to probe FLRW breakdown.  

Moreover, deviations from FLRW are expected at smaller scales, where they may impact supernovae \cite{2010MNRAS.401.1409C, Colin:2010ds, Dai:2011xm, Turnbull:2011ty, Appleby:2014kea, Colin:2018ghy, Mohayaee:2020wxf}. It is plausible that supernovae below $z \sim 0.15$ are affected by larger than expected bulk flows \cite{Colin:2010ds,Watkins:2008hf, Kashlinsky:2008ut, Feldman:2009es, Mohayaee:2020wxf, Migkas:2021zdo, Rameez:2019wdt, Migkas:2020fza}. {Interestingly, \cite{Migkas:2020fza, Migkas:2021zdo} study scaling relations in galaxy clusters for redshifts $z \lesssim 0.3$ and find in line with Figure \ref{H0LiCOW-sky} that $H_0$ varies across the sky within the flat $\Lambda$CDM model. However, the lensed quasars are much deeper in redshift, but ultimately may be pointing to the same physics.} It is also worth noting that SH0ES remove supernovae below $z \sim 0.025$, precisely because of a documented higher $H_0$ \cite{Jha:2006fm}. However, recent studies show that the higher redshift supernovae are also affected by the local bulk flow as convergence to the CMB frame at such low redshifts cannot be assumed. The effect of the bulk flow on the Hubble parameter has already been the subject of a few studies (see {\it e.g.} \cite{Hess:2014yka}). 

Overall, there are enough claims in the literature that $H_0 \sim 73$ km/s/Mpc is a compelling outcome of local $H_0$ measurements and our observation is that this central value may be offset from what any FLRW cosmology can accommodate. Different recent studies have already put the validity of FLRW in doubt (see {\it e.g.} \cite{Secrest:2020has} and references therein). Here we have argued that Hubble tension could indeed be yet another manifestation of the breakdown of the FLRW model. Alternatively, if local $H_0$ values converge lower, but still above $H_0 \sim 70$ km/s/Mpc, then we may be looking at new early Universe physics within FLRW. 

\section*{Acknowledgements} 
We thank Silvia Galli, Raphael Gavazzi, Subir Sarkar, Leandros Perivolaropoulos, Paul Steinhardt \& Kenneth Wong for correspondence and discussion. E\'OC is funded by the National Research Foundation of Korea (NRF-2020R1A2C1102899). MMShJ would like to acknowledge SarAmadan grant No. ISEF/M/400122.
LY is supported by the CQUeST of Sogang University (NRF-2020R1A6A1A03047877).

\appendix

\section{Prior on $\omega_m$}
\label{sec:prior}
We can get a prior on $\omega_{m} \equiv \Omega_{m0} h^2$ directly from \cite{Vonlanthen:2010cd}, where it is argued that one can remove the lower multipoles from the CMB to get a determination that is ``as model independent as possible" of the late-time cosmology. Excluding multipoles $\ell < 40$, while assuming that $\omega_{b}$ and $\omega_{c}$ are independent in the baryonic and cold dark matter (CDM) sectors, which will only lead to an overestimation, i. e. a conservative error, one arrives at the prior: 
\be
\label{prior1}
\omega_m = 0.145 \pm 0.007. 
\ee
This can be compared with the Planck value \cite{Aghanim:2018eyx} $\omega_{m} = 0.143 \pm 0.0011$ \cite{Aghanim:2018eyx}. Note that the error has increased by removing the lower multipoles.  

Although the data analysed in \cite{Vonlanthen:2010cd} is older (WMAP5+ACBAR), one can recast the results of \cite{Verde:2016wmz} to get a more recent prior that is valid in the Planck era. From Table 2 of \cite{Verde:2016wmz}, we have 
\bea
\Omega^{\textrm{rec}}_{b} &=& 0.1187 \pm 0.0034, \nn
\Omega^{\textrm{rec}}_{c} &=& 0.6378 \pm 0.071,  
\eea
where it should be noted that the baryonic and CDM energy densities,  $\Omega_{b}$ and $\Omega_{c}$ respectively, are evaluated at a recombination redshift of $z_{\textrm{zec}} = 1089.0 \pm 0.5$. Together with the radiation sector, they sum to unity $ \Omega_{b} + \Omega_{c} + \Omega_{r}= 1$, since the DE sector is negligible at higher redshifts. It is worth noting from Table 2 of \cite{Verde:2016wmz} that different models are considered and the energy densities for different models vary in a negligible way. This appears to back up the claim that the determinations are agnostic about the late-time cosmology. In contrast to the earlier paper \cite{Vonlanthen:2010cd}, only the $\ell < 30$ multipoles are removed. 

Once again adding the errors in quadrature, on the assumption that they are independent, one arrives at the following equality, 
\be
\frac{\omega_{m} (1+ z_{\textrm{rec}})^3}{\omega_{m} (1+ z_{\textrm{rec}})^3 + \omega_{r} (1+ z_{\textrm{rec}})^4} = 0.7565 \pm 0.0078,  \nn
\ee
once one neglects the DE sector. Here we use $\omega_{r} = 2.47 \times 10^{-5} (1+ 0.2271 N_{\textrm{eff}})$ with $N_{\textrm{eff}} = 3.046$. We further assume that errors in $z_{\textrm{rec}}$ are independent and solve for $\omega_{m}$ for a large number of configurations $\sim 10000$ subject to the assumption that the errors are Gaussian. This ultimately leads to the conservative prior: 
\be
\label{CMBprior}
\omega_m = 0.141 \pm 0.006, 
\ee
which is marginally tighter than (\ref{prior1}).

\end{document}